\documentclass[5p,times]{elsarticle}

\usepackage{amssymb,natbib,hyperref}
\usepackage{bm}
\usepackage{mathrsfs,amsfonts,amsmath}

\begin{document}
\begin{frontmatter}
\title{$\bar{D}\Sigma^*_c$ and $\bar{D}^*\Sigma_c$ interactions and the LHCb hidden-charmed pentaquarks }
\author[a,b,c]{Jun He}\ead{junhe@impcas.ac.cn}
\address[a]{Theoretical Physics Division, Institute of Modern Physics, Chinese Academy of Sciences,
Lanzhou 730000, China}
\address[b]{Research Center for Hadron and CSR Physics,
Lanzhou University and Institute of Modern Physics of CAS, Lanzhou 730000, China}
\address[c]{State Key Laboratory of Theoretical Physics, Institute of
Theoretical Physics, Chinese Academy of Sciences, Beijing  100190,China}

\begin{abstract}

Very recently, two hidden-charmed resonances $P_c(4380)$ and $P_c(4450)$ consistent with pentaquark states were observed at the LHCb detector.  The two $P_c$ states locate just below the $\bar{D}\Sigma^*_c$ and $\bar{D}^*\Sigma_c$ thresholds with mass of gaps about 5 and 15 MeV, respectively. Inspired by this fact  we perform a dynamical investigation about the $\bar{D}\Sigma_c^*(2520)$ and $\bar{D}^*\Sigma_c(2455)$ interactions which are described by the meson exchanges.  A bound state which carries spin-parity $J^P=3/2^-$ is produced from  the $\bar{D}\Sigma^*_c(2520)$ interaction, which is consistent with the $P_c(4380)$ observed at the LHCb detector. From the $D^*\Sigma_c(2455)$ interaction, a bound state with $5/2^+$ is produced, which can be related to the $P_c(4450)$. The results suggest that the $P_c(4380)$ and $P_c(4450)$ are good candidates of $\bar{D}\Sigma_c^*(2520)$ and $\bar{D}^*\Sigma_c(2455)$ molecular states, respectively.
\end{abstract}


\end{frontmatter}
\section{Introduction}

It is a long story to search for the pentaquark. In the conventional quark model, a baryon is made of three constituent quarks.  However, there exist many interpretations of baryon resonance beyond the three-quark picture. For example, the
hyperon resonance $\Lambda (1405)$ was explained as a $N\bar{K}$ bound
state by many authors~\cite{Dalitz:1960du,Oset:1997it,Jido:2010ag,Hall:2014uca}. The last rise and fall of the pentaquark study happened  ten years ago beginning with a claim of finding of the strange pentaquark $\Theta$ with mass about 1.540~GeV by
the LEPS Collaboration~\cite{Nakano:2003qx}. Much theoretical and experimental effort was put into this field and many ideas were inspired. For example, Zou $et\ al.$ proposed to include
the pentaquark components to nucleon~\cite{Zou:2005xy}. After extending to heavy flavor sector they suggested the existence of hidden charmed nucleon resonances~\cite{Wu:2010jy,Yuan:2012wz} and the discovery potential in experiment was also discussed~\cite{Wu:2009md,Huang:2013mua}. Recently, the
molecular state composed of a charmed meson and a nucleon was
discussed in the one-boson-exchange
model (OBE) ~\cite{Yasui:2009bz,He:2010zq,He:2012zd,Yamaguchi:2011xb}. Especially, the possible hidden-charm molecular baryons with the components of an anti-charmed meson and a charmed baryon were investigated systematically in Ref. \cite{Yang:2011wz}.

Two hidden-charm resonances in the $J/\psi p$ invariant mass spectrum were reported very recently by the LHCb Collaboration~\cite{arxiv-1507.03414}.  The lower one has a mass $4380\pm8\pm29$ MeV and a wider width of $205\pm18\pm86$ MeV, while the higher and narrower  one has a mass of $4449.8\pm1.7\pm2.5$ MeV and a width of $39\pm5\pm19$ MeV. The preferred $J^P$ assignments are of opposite parity, with one state having spin $3/2$ and the other $5/2$. The high mass  and  observation in the $J/\psi p$  channel suggest that these two  states must have minimal quark content of $c\bar{c} uud$, which can be called pentaquark-charmonium states.

Soon after the experimental results released, a QCD sum rule investigation was performed, by which the the $P_c(4380)$ and $P_c(4450)$  were identified as exotic hidden-charm pentaquarks composed of an anti-charmed meson and a charmed baryon \cite{Chen:2015moa}. An interpretation of the $P_c(4380)$ and $P_c(4450)$ as $S$-wave $\bar{D}^*\Sigma_c(2455)$ and $\bar{D}^*\Sigma^*_c(2520)$ molecular states were proposed in Ref. \cite{Chen:2015loa}.
A coupled-channel calculation was performed to analyze the $\Lambda_b \to J/\psi K^- p$ reaction   and supported  a $J^P=3/2^-$ assignment to $P_c(4450)$  as a molecular state mostly made of $\bar D^* \Sigma_c$ and $\bar D^* \Sigma^*_c$ \cite{Roca:2015dva}.

An important feather of these two states is that the $P_c(4380)$ and $P_c(4450)$ are very close to the $\bar{D}\Sigma_c^*(2520)$ and $\bar{D}^*\Sigma_c(2455)$ thresholds with mass gaps about 5 MeV and 15 MeV
, respectively. Considered the hadronic molecular stats is a loosely bound state, it is more natural to interpret the  $P_c(4380)$ and $P_c(4450)$ as bound states composed of $\bar{D}\Sigma_c^*(2520)$ and $\bar{D}^*\Sigma_c(2455)$ instead of $\bar{D}^*\Sigma_c(2455)$ and $\bar{D}^*\Sigma_c(2520)$ because mass gaps between observed masses and corresponding thresholds are about 80 MeV. Besides, we should not be limited to S wave which provides only negative parity state. It conflicts with the LHCb experiment which suggests the opposite parities for two $P_c$ states. In this work the  $\bar{D}\Sigma_c^*$ and $\bar{D}^*\Sigma_c$  ( numbers for mass 2520 and 2455 are omitted here and hereafter) interactions  will be studied in a Bethe-Salpeter equation approach ~\cite{He:2015mja,He:2014nya} and the bound states with both positive and negative parities  will be searched.

This work is organized as follows. After introduction, the detail of the dynamical study of  $\bar{D}\Sigma^*_c$, $\bar{D}^*\Sigma_c$ and $\bar{D}^*\Sigma^*_c$  interactions will be presented, which includes relevant effective Lagrangian and a detailed derivation of  effective potential. The numerical results  are given in Section 3. Finally, a brief summary will be given.

\section{$\bar{D}\Sigma^*_c$, $\bar{D}^*\Sigma_c$ and $\bar{D}^*\Sigma^*_c$  interactions}

In this work the interactions between  anti-charmed meson and charmed baryon will be described by the light meson exchanges.
We need to construct effective Lagrangians depicting the couplings of light mesons and  anti-charmed mesons or charmed baryons.

In terms of heavy quark limit and chiral symmetry, the couplings of light mesons to heavy-light charmed mesons $\tilde{\mathcal{P}}=(\bar{D}^0, D^-, D^-_s)$  are constructed  as \cite{Cheng:1992xi,Yan:1992gz,Wise:1992hn,Casalbuoni:1996pg},
\begin{eqnarray}\label{eq:lag-p-exch}
 \mathcal{L}_{\mathcal{\tilde{P}\tilde{P}}\mathbb{V}}
 &=&\frac{\beta{}g_V}{\sqrt{2}}\tilde{\mathcal{P}}^{\dag}_a
 \overleftrightarrow{\partial}_\mu\tilde{\mathcal{P}}_b
  \mathbb{V}^\mu_{ab}
,\label{ppv}\nonumber\\
\mathcal{L}_{\mathcal{\tilde{P}\tilde{P}}\sigma}
  &=& -2g_sm_{\mathcal{P}}\tilde{\mathcal{P}}^{}_b\tilde{\mathcal{P}}^{\dag}_b\sigma
,\nonumber\\
\mathcal{L}_{\tilde{\mathcal{P}}^*\tilde{\mathcal{P}}^*\mathbb{P}} &=&
-\frac{g}{f_\pi}\varepsilon_{\alpha\beta\lambda\kappa}\tilde{\mathcal{P}}^{*\beta\dag}_{a}
\overleftrightarrow{\partial}^\alpha{\tilde{\mathcal{P}}}^{*\kappa}_{b}
\partial^\lambda\mathbb{P}_{ab}
,\label{ppp}\nonumber\\
\mathcal{L}_{\tilde{\mathcal{P}}^*\tilde{\mathcal{P}}^*\mathbb{V}}
&=& -i\frac{\beta g_V}{\sqrt{2}}
\tilde{\mathcal{P}}^{*\dag\mu}_a\overleftrightarrow{\partial}^\nu\tilde{\mathcal{P}}_{b\mu}^{*}
\mathbb{V}_{ab\nu}\nonumber\\&&
-i2\sqrt{2}m_{\mathcal{P}^*}\lambda{}g_V\tilde{\mathcal{P}}^{*\mu\dag}_a\tilde{\mathcal{P}}^{*\nu}_b
(\partial_\mu{}
\mathbb{V}_\nu - \partial_\nu{}\mathbb{V}_\mu)_{ab}
,\nonumber\\
  \mathcal{L}_{\tilde{\mathcal{P}}^*\tilde{\mathcal{P}}^*\sigma}
  &=& 2g_sm_{\mathcal{P}^*}\tilde{\mathcal{P}}^{*}_b\cdot{}\tilde{\mathcal{P}}^{*\dag}_b\sigma
 ,\label{sps}
\end{eqnarray}
where constant $f_\pi=132$ MeV and $\mathbb
P$ and $\mathbb V$ are the pseudoscalar and vector matrices
\begin{eqnarray}
    {\mathbb P}&=&\left(\begin{array}{ccc}
        \frac{1}{\sqrt{2}}\pi^0+\frac{\eta}{\sqrt{6}}&\pi^+&K^+\\
        \pi^-&-\frac{1}{\sqrt{2}}\pi^0+\frac{\eta}{\sqrt{6}}&K^0\\
        K^-&\bar{K}^0&-\frac{2\eta}{\sqrt{6}}
\end{array}\right),\nonumber\\
\mathbb{V}&=&\left(\begin{array}{ccc}
\frac{\rho^{0}}{\sqrt{2}}+\frac{\omega}{\sqrt{2}}&\rho^{+}&K^{*+}\\
\rho^{-}&-\frac{\rho^{0}}{\sqrt{2}}+\frac{\omega}{\sqrt{2}}&K^{*0}\\
K^{*-}&\bar{K}^{*0}&\phi
\end{array}\right).
\end{eqnarray}

The effective Lagrangians depicting the charmed baryons with the light mesons with chiral symmetry, heavy quark limit and hidden local symmetry read \cite{Yan:1992gz,Liu:2011xc},
\begin{eqnarray}
{\cal L}_{\mathcal{B}_6\mathcal{B}_6\mathbb{P}}
&=&-\frac{g_1}{4f_\pi}~\epsilon^{\alpha\beta\lambda\kappa}
    \langle\bar{\mathcal{B}}_6~\overleftrightarrow{\partial}^\kappa\gamma_\alpha \gamma_\lambda
    \partial_\beta\mathbb{P} ~\mathcal{B}_6\rangle,\nonumber\\
{\cal L}_{\mathcal{B}_6\mathcal{B}_6\mathbb{V}} &=& -i\frac{\beta_S
g_{V}}{2\sqrt{2}}~\langle\bar{\mathcal{B}}_6~ \overleftrightarrow{\partial}\cdot
\mathbb{V}~ \mathcal{B}_6\rangle\nonumber\\
&&-\frac{im_{\mathcal{B}_6}\lambda_Sg_{V}}{3\sqrt{2}}~\langle\bar{\mathcal{B}}_6\gamma_\mu
\gamma_\nu
(\partial^\mu \mathbb{V}^\nu-\partial^\nu\mathbb{V}^\mu)\mathcal{B}_6\rangle,\nonumber\\
{\cal L}_{\mathcal{B}_6\mathcal{B}_6\sigma} &=&
-\ell_Sm_{\mathcal{B}_6}\langle\bar{\mathcal{B}}_6~\sigma~
\mathcal{B}_6\rangle,\label{ha1}\nonumber\\
{\cal L}_{\mathcal{B}^*_6\mathcal{B}^*_6\mathbb{P}}
&=&\frac{-3g_1}{4 f_\pi}~\epsilon^{\alpha\beta\lambda\kappa}
    \langle\bar{\mathcal{B}}^*_{6\alpha}
   \overleftrightarrow{\partial}^\kappa \partial_\beta\mathbb{P} ~\mathcal{B}^*_{6\lambda}\rangle,\nonumber\\
{\cal L}_{\mathcal{B}^*_6\mathcal{B}^*_6\mathbb{V}} &=& i\frac{\beta_S
	g_{V}}{2\sqrt{2}}~\langle\bar{\mathcal{B}}^{*\mu}_6\overleftrightarrow{\partial}\cdot
	\mathbb{V}\mathcal{B}^*_{6\mu}\rangle\nonumber\\
	&&+\frac{im_{\mathcal{B}^*_6}\lambda_Sg_{V}}{\sqrt{2}}~\langle\bar{\mathcal{B}}^{*\mu}_6
(\partial^\mu
\mathbb{V}^\nu-\partial^\nu\mathbb{V}^\mu)\mathcal{B}^*_{6\nu}\rangle,\nonumber\\
{\cal L}_{\mathcal{B}^*_6\mathcal{B}^*_6\sigma} &=&
\ell_Sm_{\mathcal{B}^*_6}\langle\bar{\mathcal{B}}^*_6~\sigma~
\mathcal{B}^*_6\rangle,\label{ha1}
\end{eqnarray}
where the partial $\overleftrightarrow{\partial}$ operates on the initial and final baryons and  $\mathcal{B}_6$ matrix is
\begin{eqnarray}
\mathcal{B}_6&=&\left(\begin{array}{ccc}
\Sigma_c^{++}&\frac{1}{\sqrt{2}}\Sigma^+_c&\frac{1}{\sqrt{2}}\Xi'^+_c\\
\frac{1}{\sqrt{2}}\Sigma^+_c&\Sigma_c^0&\frac{1}{\sqrt{2}}\Xi'^0_c\\
\frac{1}{\sqrt{2}}\Xi'^+_c&\frac{1}{\sqrt{2}}\Xi'^0_c&\Omega^0_c
\end{array}\right).
\end{eqnarray}

We list the values of the coupling constants used in the above Lagrangians in Table.
\ref{coupling}, which have been determined in literatures \cite{Liu:2011xc,Falk:1992cx,Isola:2003fh}.
The signs of the coupling constants $g$, $\beta/\lambda$ and $g_s$
are not well constrained, which will be discussed later.

\renewcommand{\arraystretch}{1.5}
\begin{table}[h!]
\caption{The parameters and coupling constants adopted in our
calculation \cite{Liu:2011xc,Falk:1992cx,Isola:2003fh}. The $\lambda$ and $\lambda_S$ are in the unit of GeV$^{-1}$. Others are in the unit of $1$.
\label{coupling}}
\begin{tabular}{cccccccccccccccccc}\hline
$\beta$&$g$&$g_V$&$\lambda$ &$g_{_S}$&$\beta_S$&$\ell_S$&$g_1$&$\lambda_S$\\
\hline
0.9&0.59&5.8&0.56 &0.76&1.74&6.2&0.94&3.31\\
\hline
\end{tabular}
\end{table}

With above Lagrangians, the potential kernels for the $\bar{D}\Sigma_c^*$  interactions from vector   and scalar meson exchanges can be written as,
\begin{eqnarray}
	{\cal V}_{\bar{D}\Sigma^*_c, \mathbb{V}}
&=& i\frac{\beta
g^2_V}{2}\left[\frac{\beta_S}{2}
	(k_2+k_2)\cdot(k_1+k'_1)\bar{\Sigma}^*_c\cdot
	\Sigma^*_c\right.\nonumber\\&-&m_{\Sigma^*_c}\lambda_S
	(\bar{\Sigma}^*\cdot q
	\Sigma^*_c\cdot(k_1+k'_1)\nonumber\\&-&\left.\bar{\Sigma}^*_c\cdot(k_1+k'_1)
~ \Sigma^*_c\cdot q\right]~P_\mathbb{V}(q^2),\nonumber\\
{\cal V}_{\bar{D}\Sigma^*_c, \sigma}
&=&i2\ell_S g_s m_Dm_{\Sigma_c^*} \Sigma_c^*\cdot \Sigma^* P_\sigma (q^2).
\end{eqnarray}
The potential kernels for the $\bar{D}^*\Sigma_c$  interactions from pseudoscalar, vector  and meson exchanges are,
\begin{eqnarray}
	{\cal V}_{\bar{D}^*\Sigma_c, \mathbb{P}}
&=& i\frac{gg_1}{4f^2_\pi}
\epsilon_{\alpha\beta\lambda\kappa}\bar{D}^{*\beta\dag}(k_1+k'_1)^\alpha
\bar{D}^{*\kappa}q^\lambda ~\epsilon^{\alpha'\beta'\lambda'\kappa'} \nonumber\\
 &\cdot&(k_2+k'_2)_{\kappa'}
 q_{\beta'}~
 \bar{\Sigma}_c~\gamma_{\alpha'} \gamma_{\lambda'}
~\Sigma_c~ P_\mathbb{V}(q^2),\nonumber\\
{\cal V}_{\bar{D}^*\Sigma_c, \mathbb{V}}
&=&i g_V^2\Big\{\frac{\beta \beta_S }{4}(k_1+k'_1)\cdot (k_2+k'_2)
\bar{D}^{*\dag}\cdot
\bar{D}^{*}~\bar{\Sigma}_c{\Sigma}\nonumber\\
&-&\frac{m_{\Sigma_c}\beta\lambda_S
}{6}[q^\mu (k_1+k'_1)^\nu-q^\nu (k_1+k'_1)^\mu]\nonumber\\
&\cdot&\bar{\Sigma}_c\gamma_\mu\gamma_\nu
\Sigma_c~\bar{D}^{*\dag}\cdot
\bar{D}^{*}+\lambda \beta_S m_{D^*}~\nonumber\\
&\cdot&
[q_\mu(k_1+k'_1)_\nu-q_\nu (k_1+k'_1)_\mu]\bar{D}^{\mu\dag}\bar{D}^{*\nu}
~\bar{\Sigma}_c\Sigma_c\nonumber\\
&-&\frac{2m_{\Sigma_c} m_{D^*}\lambda\lambda_S
}{3}\bar{\Sigma}_c[ \gamma\cdot q (q^\mu\gamma^\nu-q^\nu \gamma^\mu)\nonumber\\
&-&(q^\mu \gamma^\nu-q^\nu \gamma^\mu) \gamma\cdot q]
\Sigma_c \bar{D}^{\dag}_\mu\bar{D}^{*}_\nu\Big\}~P_\mathbb{V}(q^2),\nonumber\\
{\cal V}_{\bar{D}^*\Sigma_c^*, \sigma}&=&i2g_s \ell_Sm_{D^*}m_{\Sigma^*_c}
\bar{\Sigma}_c\Sigma_c~ \bar{D}^{*\dag}\cdot\bar{D}^*P_\sigma(q^2).
\end{eqnarray}

The potential kernels for the $\bar{D}^*\Sigma_c^*$  interactions  read,
\begin{eqnarray}%
	{\cal V}_{\bar{D}^*\Sigma_c^*, \mathbb{P}}
&=& -i\frac{3gg_1}{4f^2_\pi}
\epsilon^{\alpha\beta\lambda\kappa}\bar{D}^{*\dag}_\beta(k_1+k'_1)_\alpha
\bar{D}^{*}_{\kappa}q_\lambda \nonumber\\
&\cdot&~\epsilon^{\alpha'\beta'\lambda'\kappa'}(k_2+k'_2)_{\kappa'}
q_{\beta'}~
\bar{\Sigma}^*_{c\alpha'} \Sigma^*_{c\lambda'} ~P_\mathbb{P}(q^2),\nonumber\\
{\cal V}_{D^*\Sigma_c^*, \mathbb{V}}
&=&ig_V^2\Big\{-\frac{\beta \beta_S }{4}(k_1+k'_1)\cdot (k_2+k'_2) \nonumber\\
&\cdot&
\tilde{D}^{*\dag}\cdot
\tilde{D}^{*}\bar{\Sigma}_c^*\cdot{\Sigma}_c^*+2m_{\Sigma_c^*} m_{D^*}\lambda\lambda_S\nonumber\\&\cdot&
[\bar{D}^{*\dag}\cdot q(\bar{\Sigma}^*_c\cdot q \Sigma^*_c\cdot
	\bar{D}^*-\bar{\Sigma}^*_c\cdot \bar{D}^*\Sigma^*_c\cdot
	q)\nonumber\\
	&-&\bar{D}^*\cdot q(\bar{\Sigma}^*_c\cdot q
	\Sigma^*_c\cdot \bar{D}^{*\dag}-\bar{\Sigma}^*_c\cdot
	D^{*\dag}
\Sigma^*_c\cdot q)]\nonumber\\
&+&\frac{m_{\Sigma^*_c}\beta\lambda_S
}{2}[q^\mu (k_1+k'_1)^\nu-q^\nu (k_1+k'_1)^\mu]\nonumber\\
&\cdot&\bar{D}^{*\dag}\cdot
\bar{D}^{*}\bar{\Sigma}^{*}_{c\mu}
\Sigma^*_{c\nu}-\lambda \beta_S  m_{P^*}
[q_\mu(k_1+k'_1)_\nu\nonumber\\&-&q_\nu (k_1+k'_1)_\mu]
\bar{D}^{\mu\dag}\bar{D}^{*\nu}\bar{\Sigma}_c^*\cdot
\Sigma^*_c\Big\}P_\mathbb{V}(q^2),\nonumber\\
{\cal V}_{\bar{D}^*\Sigma_c^*, \sigma}&=&-i2g_s
\ell_Sm_{\bar{D}^*}m_{\Sigma^*_c}\bar{\Sigma}^*_c\cdot
\Sigma^*_c~\bar{D}^{*\dag}\cdot\bar{D}^* P_\sigma(q^2).\quad \
\end{eqnarray}
Here  $\bar{D}^*$, $\Sigma_c$ and $\Sigma^*_c$ mean the polarized vector, spinor and Rarita-Schwinger vector-spinor for $\bar{D}^*$ meson, and $\Sigma_c$ and $\Sigma^*_c$ baryon. The $k_{1,2}$ and $k'_{1,2}$ are the momenta for the initial and final particles with particles 1 and 2 being the anti-charmed meson and charmed baryon. The exchange momentum $q$ is defined as $q=k'_2-k_2$. The $P(q^2)$ functions for the propagator of the exchanged meson read,
\begin{eqnarray}%
P_\mathbb{P}(q^2)&=&\left(\frac{-1}{q^2-m_\pi^2}+\frac{1}{6}\frac{1}{q^2-m_\eta^2}\right),\nonumber\\
P_\mathbb{V}(q^2)&=&\left(\frac{-1}{q^2-m_\rho^2}-\frac{1}{2}\frac{1}{q^2-m_\omega^2}\right),\nonumber\\
P_\sigma(q^2)&=&\frac{1}{q^2-m^2_\sigma}.
\end{eqnarray}
In our previous work~\cite{He:2015mja}, the $J/\psi$ exchange was included for the $DD^*$ interaction because of the OZI suppression of light meson exchange. In this work, there does not exist such OZI suppression, so we assume the light meson is dominant with an argument that the exchange from a heavy-mass meson should be suppressed. Besides, the coupled-channel effect is assumed to be small as usual OBE model~\cite{Yang:2011wz,Chen:2015moa}.

In this work we will adopt a Bethe-Saltpeter approach with a spectator quasipotential approximation, which was explained explicitly in the appendices of Ref.~\cite{He:2015mja}, to search the possible bound states related to the $P_c$ states. The bound state (or resonance) from the interaction corresponds to the pole of the scattering amplitude ${\cal M}$  which is described by potential kernel obtained in the above. The Bethe-Saltpeter equation for partial-wave amplitude with fixed spin-parity $J^P$ reads ~\cite{He:2015mja},
\begin{eqnarray}
{\cal M}^{J^P}_{\lambda'\lambda}({\rm p}',{\rm p})
&=&{\cal V}^{J^P}_{\lambda',\lambda}({\rm p}',{\rm
p})+\sum_{\lambda''}\int\frac{{\rm
p}''^2d{\rm p}''}{(2\pi)^3}\nonumber\\
&\cdot&
{\cal V}^{J^P}_{\lambda'\lambda''}({\rm p}',{\rm p}'')
G_0({\rm p}''){\cal M}^{J^P}_{\lambda''\lambda}({\rm p}'',{\rm
p}).\quad\quad \label{Eq: BS_PWA}
\end{eqnarray}
Note that the sum extends only over nonnegative helicity $\lambda''$.
The partial wave potential is defined as
\begin{eqnarray}
{\cal V}_{\lambda'\lambda}^{J^P}({\rm p}',{\rm p})
&=&2\pi\int d\cos\theta
~[d^{J}_{\lambda\lambda'}(\theta)
{\cal V}_{\lambda'\lambda}({\bm p}',{\bm p})\nonumber\\
&+&\eta d^{J}_{-\lambda\lambda'}(\theta)
{\cal V}_{\lambda'-\lambda}({\bm p}',{\bm p})],
\end{eqnarray}
where the initial and final relative momenta are chosen as ${\bm p}=(0,0,{\rm p})$  and ${\bm p}'=({\rm p}'\sin\theta,0,{\rm p}'\cos\theta)$ with ${\rm p}^{(')}=|{\bm p}^{(')}|$. The $d^J_{\lambda\lambda'}(\theta)$ is the Wigner d-matrix.  We would like to note that the partial wave decomposition here is done into the quantum number $J^P$ instead of usual orbital angular momentum $L$ so that all partial waves based on $L$ related to a certain $J^P$ considered  are included. It is an advantage of our method because the experiment result is usually provided with spin parity $J^P$.

 In this work we will adopt an  exponential
regularization by introducing a form factor in the propagator as
\begin{eqnarray}
	G_0({\rm p})\to G_0({\rm p})\left[e^{-(k_1^2-m_1^2)^2/\Lambda^4}\right]^2,\label{Eq: FFG}
\end{eqnarray}
with $k_1$ and $m_1$ being the momentum and mass of the charmed meson. The interested reader is referred to Ref.~\cite{He:2015mja} for further information about the regularization.
 A form factor is introduced to compensate the
off-shell effect of exchange meson  as  $f(q^2)=(\frac{\Lambda^2}{\Lambda^2-q^2})^4$. Considered the similar roles played by two cut offs, the same value  for two cut offs will be adopted for simplification.

\section{Numerical results}

After transforming the integral equation to a matrix equation, the pole of scattering  amplitude $\cal M$ can be searched by variation of $z$ to satisfy
$|1-V(z)G(z)|=0$
with  $z=M+i\Gamma/2$ equaling to the system energy $M$ at the
real axis~\cite{He:2015mja}.  In this work, we will vary the cut off $\Lambda$ from 0.5
to 5 GeV to search for the poles which correspond to the bound states
or resonances from the interactions. For the sake of completeness, the dependence
of the system energy $M$ on $\Lambda$ under different combinations of
the signs of the coupling constants $g$, $\beta/\lambda$ and $g_s$ will be presented. In this work, only states with  half isospin,which are related to the $P_c$ state observed at LHCb, will be considered.

\subsection{Bound states from the $\bar{D}\Sigma^*_c$ interaction}

First, we consider the $\bar{D}\Sigma^*_c$ interaction which threshold is a
little above the lower $P_c$ observed in LHCb. Here the total
spin-parity $J^P=3/2^\pm$ is considered and only states with  mass larger
than 4.32 GeV are presented in Fig.~\ref{Fig:
DSigmaA}.

\begin{figure}[h!]
\begin{center}
\includegraphics[bb=15 10 730 570,clip, scale=0.33]{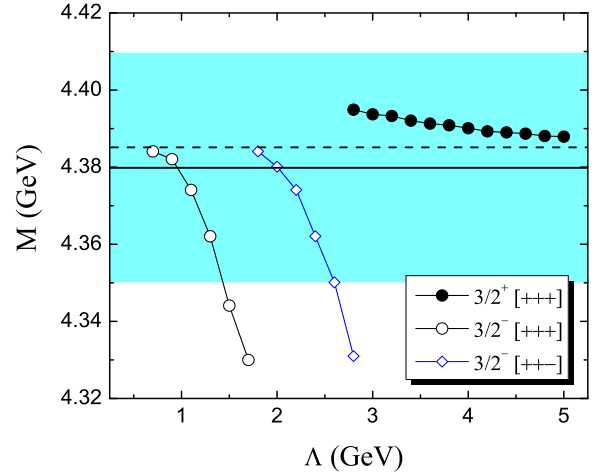}
\caption{The cut off $\Lambda$ dependence of the mass $M$ of the $\bar{D}\Sigma^*_c$ system with different spin parities $J^P$.
The solid line and band are for the experimental mass  and the uncertainty
of the $P_c(4380)$. The dashed line is for the $\bar{D}\Sigma_c^*$ threshold.
Here, $+/-$ in "$\pm\pm\pm$" denotes that we need to multiply the
corresponding sigma, vector and pion exchange potentials by an extra
factor, +1/-1, which comes from the changes of the signs of the coupling constants.
\label{Fig: DSigmaA}}
\end{center}
\end{figure}

A bound state is found at cut off about 1 GeV for the case with
quantum number $3/2^-$, which is the suggested spin parity of the $P_c(4380)$.
The bound state appears near threshold at cut of about 0.7 GeV and the
total mass of the system decreases with increase of the cut off and reaches the experimental central value of the $P_c(4380)$ mass at cut off
about 1 GeV. Although the dependence of results on the cut off is weaker than the non-relativistic OBE model ~\cite{Yang:2011wz,Chen:2015moa}, the results here are still sensitive to the cut off, which is inherited from the  non-relativistic OBE model.  It may suggest that there is still something deficient in the current model, which is absorbed in the free parameter, cut off $\Lambda$.

For the case with $3/2^+$, no bound state below the
threshold can be found while a resonance appears at cut off about 3 GeV
and approaches to the threshold slowly with increase of cut off. Its mass is above the
threshold but still within the
experimental uncertainty.  Except the sign
combinations $+++$ and $++-$, no poles are found with cut off in a range of form 0.5 to 5 GeV.

\subsection{Bound states from the $\bar{D}^*\Sigma_c$ interaction}

\begin{figure}[h!]
\begin{center}
\includegraphics[bb=15 20 730 570,clip,	scale=0.335]{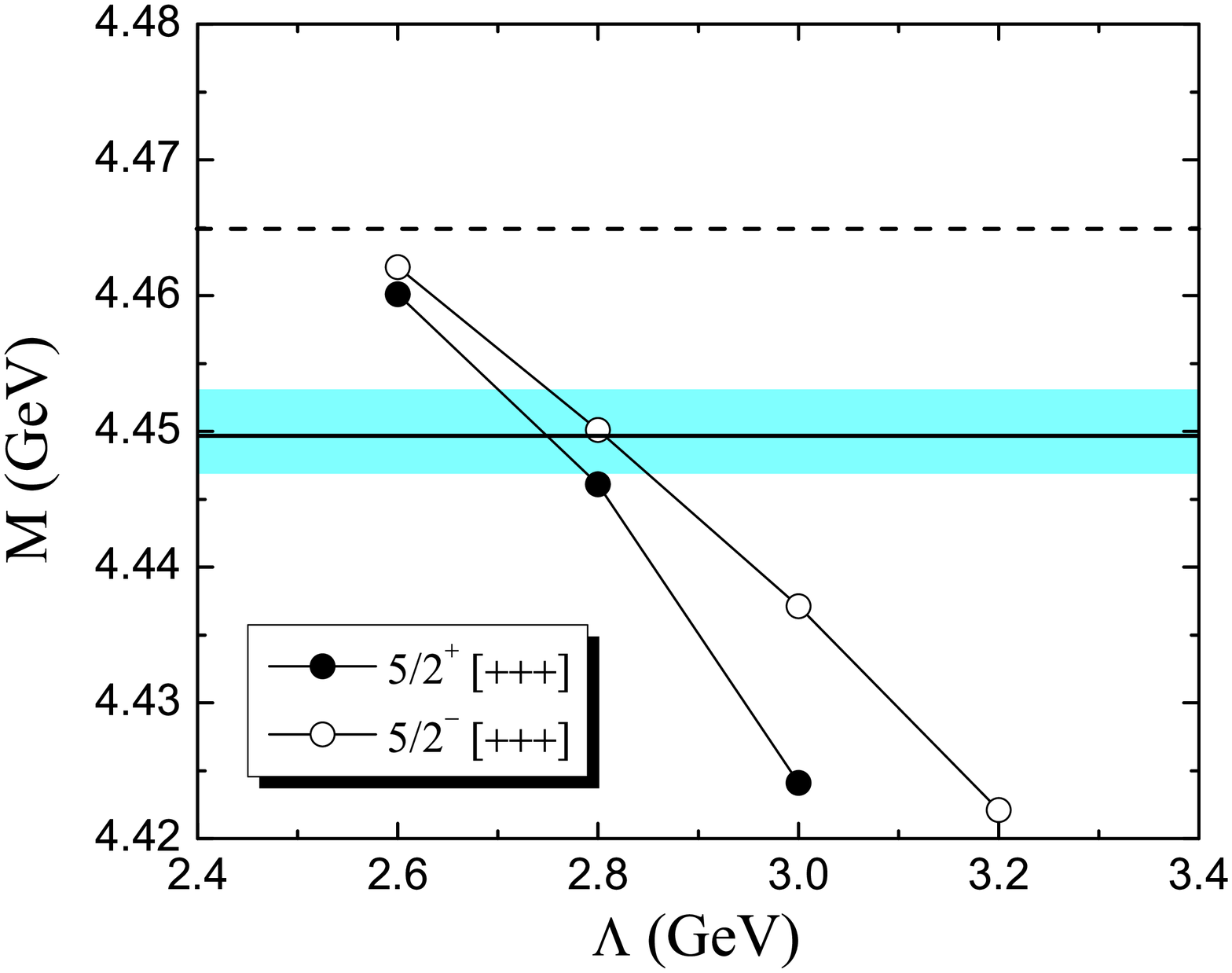}\\
\includegraphics[bb=15 20 730 570,clip, scale=0.335]{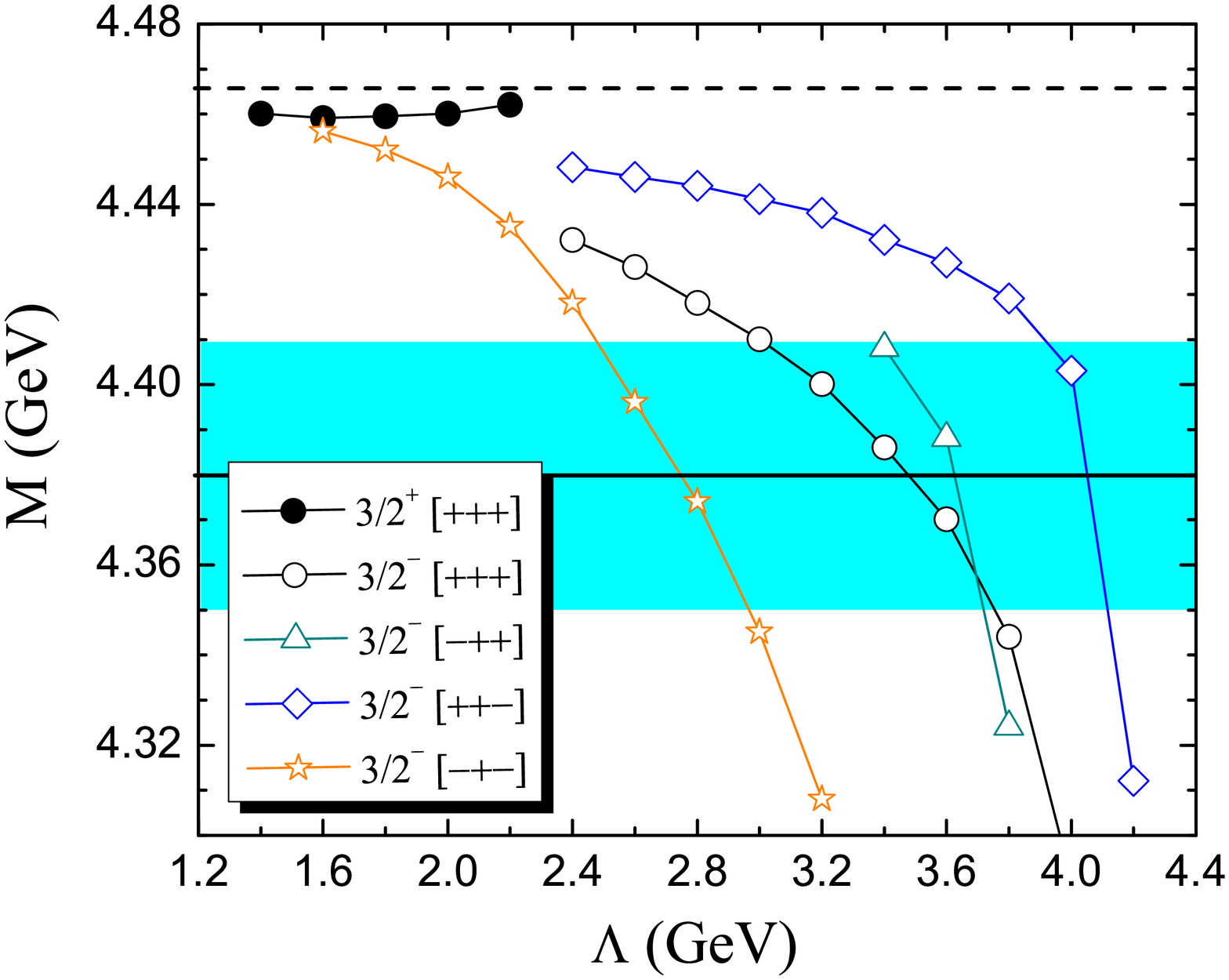}
\caption{The cut off $\Lambda$ dependence of the mass $M$ of the $\bar{D}^*\Sigma_c$ system with different spin parities $J^P$.
The upper and lower figures are for spins $J=3/2$ and
$5/2$, respectively. The solid line and band are for the experimental
mass and the uncertainty of the $P_c(4450)$ (upper figure) or
the $P_c(4380)$ (lower figure). The dashed line is for the $\bar{D}^*\Sigma_c$
threshold. The sign combination $[\pm\pm\pm]$ is the same as in Fig. \ref{Fig: DSigmaA}. \label{Fig: DASigma} } \end{center} \end{figure}

The second $P_c$ observed at LHCb locates just below the
$\bar{D}^*\Sigma_c$ threshold with a mass gap about 15 MeV. So it is natural
to relate the $P_c(4450)$ to the $\bar{D}^*\Sigma_c$ interaction. In the upper
figure of Fig. \ref{Fig: DASigma}, the $\Lambda$ dependence of the bound states from the $\bar{D}^*\Sigma_c$ interaction with
$5/2$ are presented. The poles can be produced  only with the sign combination
$+++$. The bound states appears at cut off about 2.6 GeV and then the system masses decrease to about 4.42 GeV at 3.0 and 3.2 GeV for positive and negative parities, respectively.
 The masses of the bound states reach 4.45 GeV  at cut off about 2.8 GeV.

In Ref. \cite{Chen:2015loa}, the  $P_c(4380)$ was interpreted
as a $\bar{D}^*\Sigma_c$ molecular state with a binding energy about 80 GeV. Here the results
with $3/2$ is also presented for completeness in the lower figure of Fig. \ref{Fig: DASigma}
and compared with the LHCb experiment . A bound state with
$3/2^-$ can be found  at cut off about 2.4 GeV and the mass decreases to the
experimental value with increase of cut off and reach 4.38 GeV at
cut off about 3.5 GeV.  A bound state near threshold can be found at cut off about 1.5 GeV with a positive parity.
With sign combinations $[-++]$, [++-] and [-+-], bound states are also produced from the  $\bar{D}^*\Sigma_c$ interaction.

\subsection{Bound states from the $\bar{D}^*\Sigma^*_c$ interaction}

\begin{figure}[h!]
\begin{center}
\includegraphics[bb=15 20 730 570,clip, scale=0.337]{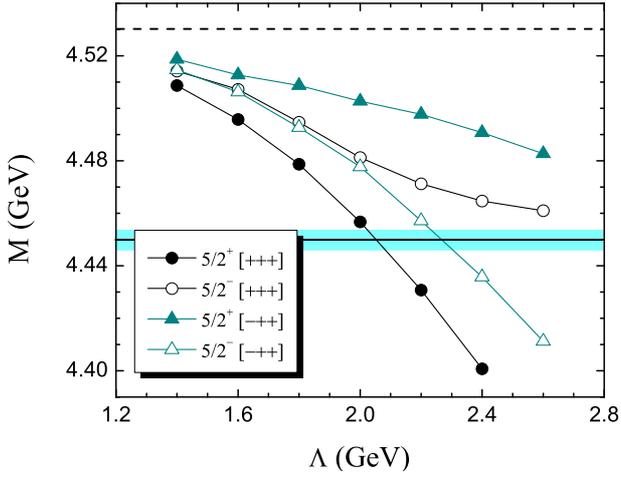}
\caption{The cut off $\Lambda$ dependence of the mass $M$ of the $\bar{D}^*\Sigma^*_c$ system with different spin parities $J^P$. The solid line and band are for the experimental mass  and the uncertainty of the $P_c(4450)$. The dashed line is for the $\bar{D}^*\Sigma^*_c$ threshold. The sign combination $[\pm\pm\pm]$ is the same as in Fig. \ref{Fig: DSigmaA}.}
\end{center}
\label{Fig: DASigmaA}
\end{figure}

Through the $\bar{D}^*\Sigma^*_c$ threshold is higher than the
 $P_c(4450)$ about 70 MeV, the results are presented for completeness. The results with $J=5/2$
are presented in Fig~\ref{Fig: DASigmaA} and compared with the
experimental mass of the $P_c(4450)$.

Bound states with  both positive and negative parities are produced from the $\bar{D}^*\Sigma^*_c$ interaction. At cut off about 2 GeV, a
bound state with the experimental mass of $P_c(4450)$ can be found
with positive parities. However, for negative parity, the pole
disappears before the mass reach the experimental value. With sign combination $[-++]$, bound states for both parities are found.

The importance of the vector meson exchanges can be seen obviously from
the results under different sign combinations. All poles are found with
the sign assignment for vector meson exchange is $+$.
Since the two experimental observed $P_c$ states can be reproduced
simultaneously only with sign combination $[+++]$, all discussions  below will be based on this sign combination. Such sign combination  is consistent with the SU(4) Lagrangians~\cite{Nagahiro:2008cv}.

\subsection{Identification of the $P_c(4380)$ and $P_c(4450)$}

We collect the possible
internal structures of the $P_c(4380)$ and $P_c(4450)$ based on the masses  as below

\begin{flushleft}
\begin{tabular}{lll}
$P_c(4380)$:& $\bar{D}\Sigma_c^* \ [3/2^-, 0.7{\rm-}1.4]$,& $\bar{D}\Sigma_c^*\ [3/2^+, 2.8{\rm -}5.0]$, \nonumber\\
& $\bar{D}^*\Sigma_c [3/2^-, 3.0{\rm-}3.7]$; \nonumber\\
$P_c(4450)$:& $\bar{D}^*\Sigma_c [5/2^+, 2.7{\rm-}2.8]$,& $\bar{D}^*\Sigma_c [5/2^-, 2.75{\rm -}2.85]$, \nonumber\\
& $\bar{D}^*\Sigma^*_c [5/2^+, 2{\rm-}2.1]$.  \nonumber
\end{tabular}
\end{flushleft}
The values in the bracket are spin-parity of the system and the cut
offs in the unit of GeV which produces the experimental mass within uncertainties.

Empirically the pole far away from the threshold provides much smaller contribution than the one near threshold. It can be also seen from the form factor in Eq.~(\ref{Eq: FFG}) which suppresses the contribution with two particles being off-shell. Hence, the $\bar{D}^*\Sigma_c[3/2^-]$ and $\bar{D}^*\Sigma^*_c[5/2^+]$ are not preferred to interpret the LHCb hidden-charmed states.
The LHCb Collaboration stated the best fit combination finds two $P_c$ states with $J^P$ values of $3/2^-$ and $5/2^+$ \cite{arxiv-1507.03414}.  Hence, we can identify the $P_c(4380)$ and $P_c(4450)$ as
\begin{eqnarray}
P_c(4380): \bar{D}\Sigma_c^* [3/2^-];\quad
P_c(4450): \bar{D}^*\Sigma_c [5/2^+].  \nonumber
\end{eqnarray}
The cut off to produce the experimental mass of the $P_c(4380)$ is about 1 GeV. The positive parity state $\bar{D}\Sigma_c^* [3/2^+]$ vanishes at such cut off. The negative parity state  $\bar{D}^*\Sigma_c(5/2^-)$ may also contribute to the $P_c(4450)$ due to the similar cut off. However, the positive parity state is in $P$ and $F$ waves while the negative parity state is in $D$ and $G$  waves so that the contribution from the latter state will be suppressed.

\section{Summary}

The  hidden-charm states $P_c(4380)$ and $P_c(4450)$ newly observed at LHCb locate just below the $\bar{D}\Sigma_c^*$ and $\bar{D}^*\Sigma_c$ thresholds, which suggests that theses states are from the $\bar{D}\Sigma_c^*$ and $\bar{D}^*\Sigma_c$ interactions. The exchanges from pseudoscalar, vector and scalar meson exchanges are included to describe the interactions of the anti-charmed meson and charmed baryon. It is found that the vector meson plays important roles in the interactions considered in this work. A Bethe-Saltpeter equation approach is applied  to find the bound state from the $\bar{D}\Sigma_c^*$ and $\bar{D}^*\Sigma_c$ interactions by searching the poles from the scattering amplitude.  The observed  $P_c(4380)$ and $P_c(4450)$ can be identified as loosely bound states, that is, hadronic molecular states, composed of $\bar{D}\Sigma_c^*$ and $\bar{D}^*\Sigma_c$ with the spin parity $J^P=3/2^-$ and $5/2^+$, respectively.

\section*{Acknowledgements}
This project is supported by the Major
State Basic Research Development Program in China
under Grant 2014CB845405 and the National Natural Science Foundation of China under Grant 11275235.

\bibliographystyle{elsarticle-num}

\end{document}